\documentclass[runningheads]{llncs}
\usepackage[T1]{fontenc}
\usepackage{graphicx}
\usepackage{enumitem}
\usepackage{booktabs}
\usepackage{multirow}
\usepackage{tabularx}
\usepackage{makecell}
\usepackage{amsmath} 
\usepackage{comment}
\usepackage{hyperref}
\usepackage{url}

\begin{document}

\title{Evaluating Argon2 Adoption and Effectiveness in Real-World Software}

\author{Pascal Tippe\inst{1}\thanks{Corresponding author: firstname.lastname@fernuni-hagen.de} \and
Michael P. Berner\inst{1}}
\institute{FernUniversität in Hagen \\ Hagen, Germany}

\maketitle          

\begin{abstract}
    Modern password hashing remains a critical defense against credential cracking, yet the transition from theoretically secure algorithms to robust real-world implementations remains fraught with challenges. This paper presents a dual analysis of Argon2, the Password Hashing Competition winner, combining attack simulations quantifying how parameter configurations impact guessing costs under realistic budgets, with the first large-scale empirical study of Argon2 adoption across public GitHub software repositories. Our economic model, validated against cryptocurrency mining benchmarks, demonstrates that OWASP's recommended 46 MiB configuration reduces compromise rates by 42.5\% compared to SHA-256 at \$1/account attack budgets for strong user passwords. However, memory-hardness exhibits diminishing returns as increasing allocations to RFC 9106's 2048 MiB provides just 23.3\% (\$1) and 17.7\% (\$20) additional protection despite 44.5\texttimes greater memory demands. Crucially, both configurations fail to mitigate risks from weak passwords, with 96.9-99.8\% compromise rates for RockYou-like credentials regardless of algorithm choice. Our repository analysis shows accelerating Argon2 adoption, yet weak configuration practices: 46.6\% of deployments use weaker-than-OWASP parameters. Surprisingly, sensitive applications (password managers, encryption tools) show no stronger configurations than general software. Our findings highlight that a secure algorithm alone cannot ensure security, effective parameter guidance and developer education remain essential for realizing Argon2's theoretical advantages.
\end{abstract}

\section{Introduction}

As reliance on digital systems continues to grow, ensuring secure user authentication has become a critical challenge in cybersecurity. Password hashing functions play a pivotal role in protecting credentials, yet legacy algorithms like SHA-256 exhibit persistent vulnerabilities when deployed in authentication systems. Despite their theoretical security properties, these algorithms are increasingly ineffective against modern attacks that leverage GPU and ASIC-based hardware to test billions of password candidates per second. This gap between theoretical robustness and practical resilience underscores the need for more advanced cryptographic solutions. Argon2, the winner of the 2015 Password Hashing Competition, represents a significant advancement in password hashing design by introducing memory-hardness that increases computational costs for attackers using specialized cracking hardware \cite{rfc9106}. By requiring substantial RAM allocation during hash computation (configurable via parameters), Argon2 creates asymmetric costs that favor defenders over attackers. However, its effectiveness depends heavily on parameter selection, with configurations such as the OWASP-recommended 46 MiB memory differing significantly from the RFC 9106 proposal of 2 GiB memory. This disparity highlights the importance of standardized, context-aware parameterization frameworks that balance security and performance. While Argon2's theoretical advantages are well-documented in academic literature, its adoption in real-world software remains inconsistent. Our preliminary repository scans suggest a persistent reliance on older algorithms like SHA-256 and PBKDF2, particularly in legacy systems where backward compatibility concerns often outweigh security considerations.  Even when Argon2 is implemented, suboptimal parameter configurations are common, reflecting systemic barriers to cryptographic modernization. This study addresses four key questions: (1) How do specific Argon2 parameter configurations compare to SHA-256 in resisting GPU/ASIC-accelerated attacks under realistic password strength assumptions? (2) What is the current state of Argon2 adoption among software projects? (3) How have Argon2 parameters evolved over time to adapt to increasing computational capabilities of potential attackers? (4) Do applications with heightened security requirements, such as password managers or file encryption software, tend to implement stronger parameter configurations?

The remainder of this paper is organized as follows: Section \ref{sec:background} reviews related work on password hashing functions and guessing attacks. Section \ref{sec:methodology} describes our methodology, while Section \ref{sec:AttackSimFramework} details our attack simulations. Section \ref{sec:AttackSimResults} presents an analysis of offline password cracking results, followed by Section \ref{sec:data-collection}'s exploration of real-world data collected from software repositories. Finally, Section \ref{sec:repo-analysis} analyses the gathered data and Section \ref{sec:Discussion} discusses findings before concluding in Section \ref{sec:Conclusion}.
\section{Background and Related Work}\label{sec:background}

Online platforms continuously suffer from breaches exposing user passwords en masse. To mitigate this risk, traditional password storage relies on cryptographic hashing to prevent credential exposure. Unlike encryption, deterministic hash functions like SHA-256 produce fixed-length digests that cannot be feasibly reversed. However, attackers can enumerate password candidates until they find a matching hash value. This attack vector fundamentally depends on two factors: the computational efficiency of the hash function and the statistical distribution of password \textit{guessability} across user populations. To prevent attackers from using precomputed rainbow tables and attacking multiple breached credentials simultaneously \cite{10.1007/978-3-540-45146-4_36}, defenders add a randomly generated string (called salt) to passwords, forcing attackers to target each individual salted password hash. While salting prevents batch attacks, it does not address the fundamental vulnerability of fast hash algorithms to brute-force and dictionary attacks. Instead of exhaustively iterating over all possible character combinations, attackers exploit users' tendency to follow predictable patterns during password creation: combining words, replacing characters with numbers, or appending special symbols. Deprecated password metrics like Shannon entropy fail to capture these human-chosen password patterns, leading researchers to develop more accurate estimation techniques. The zxcvbn algorithm's \cite{10.5555/3241094.3241108} pattern-aware entropy and models like Markov chains \cite{10.1109/SP.2009.8} or context-free grammars \cite{10.1109/SP.2009.8} better reflect real-world password weaknesses by analyzing dictionary matches, spatial keyboard patterns, and breach recurrence patterns. Bonneau \cite{6234435} formalized the concept of guessability, establishing a direct connection to practical password strength measurement by quantifying the average number of guesses required to breach a target percentage of user accounts. This approach better captures systematic risks since Dell'Amico et al. \cite{10.5555/1833515.1833671} determined an attacker would require 149,053,078 attempts on average to crack over half the passwords in three real-world datasets (approximately $2^{27.14}$). Bonneau \cite{6234435} calculated median attack costs ranging between $2^{19.8}$ and $2^{21.6}$ attempts across multiple leaked datasets including RockYou to crack 50\% of the passwords. Florencio and Herley \cite{10.1145/1242572.1242661} found an average bit-strength of approximately 40.54 bits using naive calculations across half a million users over three months.

Memory-hard functions represent a paradigm shift in password hashing by imposing substantial RAM requirements. The interplay between memory-hardness and guessability becomes apparent in Blocki et al.'s framework~\cite{8418642}, where attackers maximize compromised accounts within fixed budgets. The objective is not to linearly increase costs for both defenders and attackers by iterating hash functions, but to create asymmetric costs that disproportionately disadvantage attackers using specialized hardware such as ASICs. Argon2, the Password Hashing Competition winner in 2015, implements this approach through three tunable parameters: memory cost (\textit{m}), iterations (\textit{t}), and parallelism (\textit{p}). Argon2 comes in three variants: Argon2d (fast but vulnerable to side-channels), Argon2i (side-channel resistant but slower), and Argon2id (a hybrid approach). This study focuses exclusively on Argon2id, which RFC 9106 recommends as the default for password hashing. Its design prioritizes time-memory tradeoff resistance, forcing attackers to spend either prohibitive time or memory resources, directly impacting guessability economics. Blocki et al. \cite{8418642} modeled this increased strength against guessing attacks as an optimization problem, using Bitcoin mining hardware and blockchain hashrate as proxies for determining attacker strength, finding that memory-hard hash functions substantially increase guessing costs. Argon2's security relies on selecting appropriate values for its parameters, yet many developers in user studies struggle to implement even basic password hashing correctly \cite{10.1145/3290605.3300370}. Furthermore, while OWASP recommends 46 MiB of memory for general use cases \cite{owasp2023argon2}, the RFC 9106 standard \cite{rfc9106} advocates for significantly higher values (2048 MiB). This 44.5\texttimes difference reflects a tension between practical deployment constraints and theoretical security requirements, potentially leaving developers uncertain about which configuration is suitable for their specific applications. Our work bridges this gap by analyzing real-world Argon2 implementation patterns across GitHub repositories, quantifying the security impact of different parameter configurations through attack simulations, and identifying systemic mismatches between academic parameter recommendations and developer implementation practices. This research extends prior work by providing concrete evidence of how theoretical security advantages translate, or fail to translate, into practical security improvements in deployed software.
\section{Methodology}\label{sec:methodology}
This study employs a comprehensive methodology to evaluate the technical performance and real-world adoption of Argon2 as a password hashing algorithm. The analysis is divided into two interconnected components: a security analysis of Argon2 configurations compared to SHA-256 and an empirical investigation into the adoption trends of Argon2 across software repositories on GitHub. By combining cryptographic modeling, password strength estimation, attack simulation, and repository analysis, this methodology provides a holistic view of both theoretical efficacy and practical implementation.

\subsection{Security Analysis Framework}

The security analysis focuses on Argon2's resistance to offline password cracking under realistic attacker constraints. The threat model assumes an attacker with offline access to hashed credentials and computational resources comparable to large-scale cryptocurrency mining operations. Attackers are assumed to prioritize cost-efficiency, spending a fixed budget for cracking passwords. We leverage public password datasets for strength estimation. To model cryptographic costs, we analyze the economics of cryptocurrency mining as a proxy for attacker resources. Bitcoin's SHA-256 implementation serves as the baseline for traditional hashing costs, while Monero's RandomX, a memory-hard proof-of-work algorithm based on Argon2d, provides insights into memory-dependent computation costs. These models are validated using energy consumption benchmarks from consumer-grade CPUs, ensuring real-world applicability. Password strength estimation is conducted using zxcvbn's pattern-aware entropy metric. The RockYou 2009 dataset, leaked cleartext passwords from an online platform, is used as the baseline for password distributions, filtered to include only passwords meeting modern length requirements ($\geq$8 characters). For modeling enhanced password policies, we generated a synthetic dataset by doubling the zxcvbn bit-strength values from the filtered RockYou data. This transformation simulates passwords with significantly higher guessing resistance (e.g., a 20-bit password becomes a 40-bit password) while preserving the overall distribution characteristics. Attack simulations evaluate the effectiveness of different hashing configurations under three budget scenarios for attackers: \$0.1, \$1, and \$20 per targeted account. Analyzed configurations include SHA-256 as a baseline and Argon2 implementations with both RFC 9106 recommended parameters (2048 MiB memory) and OWASP-suggested hardened parameters (46 MiB memory).

\subsection{Data Collection and Repository Analysis}

To assess Argon2's adoption in real-world software projects, we systematically collect data from public repositories on GitHub using its REST API. GitHub was chosen due to its prominence in open-source development and its extensive repository metadata, which includes indicators such as stars that we use as a proxy for repository quality. While acknowledging that user motivations for starring repositories vary, prior research suggests that stars are more reliable indicators of relevance than other metrics like number of forks \cite{BORGES2018112}. The analysis employs two complementary search methods: repository metadata search and code search. 
Repository searches query titles, descriptions, and topics for keywords related to password hashing algorithms (\textit{Argon2}, \textit{bcrypt}, \textit{scrypt}, \textit{yescrypt} and \textit{PBKDF2}). The selection was driven by their prominence as widely recognized password hashing algorithms, providing a comparative baseline to evaluate Argon2's adoption and security properties against established standards with distinct characteristics in memory-hardness and performance. Since GitHub limits search results to 1,000 entries per query, searches are segmented by repository creation date to capture a comprehensive dataset. Code searches identify instances of password hashing algorithm implementations within source code files. To address GitHub's indexing limitations for code searches, results are segmented by programming language. Languages were selected based on their support for symbol extraction on GitHub and manual reviews of preliminary data\footnote{Selected languages: Bash, C, C\#, C++, CodeQL, Dart, Elixir, Erlang, Go, Haskell, Java, JavaScript, Kotlin, Lua, PHP, Python, R, Ruby, Rust, Scala, Starlak, Swift, TypeScript}. To ensure accuracy in both search methods, filtering mechanisms are applied to exclude false positives (e.g., repositories unrelated to password hashing or those associated with cryptocurrencies). Automated exclusion based on keywords is supplemented by manual refinement to further reduce noise in the dataset.

\subsection{Manual Review and Parameter Analysis}\label{subsec:meth-parameter}

Repositories identified through searches undergo manual review to extract Argon2 parameter configurations and classify software types. This step ensures accuracy by accounting for variations in parameter naming conventions and library usage that automated tools might miss or misclassify. Additionally, this process verifies that Argon2id is used appropriately within repositories and not in contexts such as cryptocurrency mining. Repositories where parameter configurations cannot be assessed or that serve non-productive purposes (e.g., specifications or benchmarking tools) are excluded from further analysis. To focus on high-quality implementations, only repositories with significant number of stars are included in the final dataset. The extracted parameter configurations are analyzed to evaluate their alignment with recommended security practices. Repositories are categorized by software type (e.g., web applications, password managers), allowing comparisons between parameter strengths across different application domains. To analyze trends in Argon2 adoption over time and across software categories, statistical hypothesis testing is employed. Non-parametric tests such as chi-square goodness-of-fit and independence tests examine whether observed distributions deviate significantly from uniformity or exhibit associations between variables (e.g., repository type and parameter strength). A significance level of p=0.05 is used throughout the analysis.
\section{Attack Simulation Framework}\label{sec:AttackSimFramework}

The attack simulation framework evaluates Argon2's economic resistance to offline password guessing by modeling adversarial cost structures under realistic resource constraints. Our analysis compares two recommended parameter configurations representing different security philosophies: the RFC 9106 recommendation (2048 MiB memory) prioritizing ASIC resistance through substantial memory demands, and OWASP's pragmatic guidelines (46 MiB memory) balancing security with server resource limitations. These configurations create a 44.5\texttimes difference in memory allocation, enabling direct comparison in thwarting large-scale attacks.

\subsection{Parameter Configurations}

To explore the trade-offs between security and resource efficiency, we analyze two widely referenced Argon2 parameter configurations: the RFC 9106 recommendation and OWASP's pragmatic guidelines. The RFC 9106 configuration prioritizes resistance to attacks by allocating 2048 MiB of memory per hash computation, thereby imposing significant memory demands on attackers and defenders. In contrast, OWASP's configuration uses a reduced memory allocation of 46 MiB, reflecting a balance between security and server-side performance constraints. These configurations represent distinct security philosophies, with the former emphasizing robustness against specialized hardware and the latter accommodating practical deployment scenarios. Both are the first recommended configuration and use parameters $t=1$ and $p=1$ allowing a direct comparison. The 44.5\texttimes difference in memory allocation between these configurations provides a valuable basis for evaluating their relative effectiveness in thwarting large-scale attacks. In our simulations, attackers are modeled as having fixed budgets of \$0.10, \$1.00, and \$20.00 per targeted account. These budgets reflect varying levels of attacker investment, from low-cost opportunistic attacks to more resource-intensive campaigns targeting higher value accounts. The budgetary constraints are used to calculate the number of hash computations an attacker can afford under each parameter configuration, enabling direct comparisons of their economic resistance.

\subsection{Cost per Hash Evaluation}

The computational cost of Argon2 is central to its ability to resist offline attacks. To estimate this cost, we use cryptocurrency mining as a proxy for adversarial resource expenditures due to its well-documented economic metrics and operational similarities to password cracking. Specifically, we derive baseline costs for SHA-256 from Bitcoin mining data and extrapolate Argon2 costs using Monero's RandomX algorithm, which incorporates Argon2d to create an initial cache and extends it with additional computations inside a virtual machine. For SHA-256, Bitcoin's current network hashrate (701.72 EH/s) and block rewards as of 20 February 2025 \cite{bitcoin2025stats} provide a per-hash cost estimate of approximately $\$7.079\times10^{-19}$. Argon2's memory-hardness complicates direct benchmarking. However, RandomX \cite{tevadorrandomx} serves as a functional analog due to its use of approximately 2 GiB memory allocations and Argon2d usage as a base element. Adjusting for RandomX's additional computational overhead (conservatively estimated at 100\texttimes), we estimate Argon2's base cost at $\$2.729\times10^{-12}$ per hash for 2 GiB configurations with the network statistics on 20. February 2025 (4.54 GH/s, 32 blocks per hour and 232.31\$ per unit) \cite{bitinfocharts2025monero}. This cost scales linearly with reduced memory allocations, allowing us to model the economic impact of different parameter settings.

To validate these estimates, we conducted energy consumption calculations using processor thermal design power (TDP) values and measured hashes per second on consumer-grade CPUs\footnote{Intel Core i3-7130U, AMD FX-6300, Intel Core i5-10300H, Intel Core i5-9400F, AMD Ryzen 5 2600X}. For example, using an energy price of \$0.05/kWh and considering only CPU power consumption, the cost per hash was calculated as $\$4.17 \times 10^{-7}$, which exceeds our baseline estimate derived from RandomX mining data. This discrepancy underscores the pessimistic nature of our baseline assumptions but also highlights the real-world feasibility of our cost model for very resourceful attackers.

\subsection{Dataset Preparation}

The datasets used in this study are critical for simulating realistic attack scenarios and evaluating password strength distributions under different hashing configurations. We employ two datasets: the RockYou dataset and a synthetic dataset $D_{syn}$ derived from it. The RockYou dataset, leaked in 2009, contains over 32.6 million user passwords and is an important resource in password security research due to its size and real-world origins \cite{6234435,8418642}. To ensure consistency and relevance to contemporary minimal security standards, we preprocessed this dataset. First, all passwords were normalized to UTF-8 encoding using Python scripts equipped with the \texttt{chardet} library to resolve character encoding inconsistencies; entries with unresolvable issues were removed (affecting 242 passwords). Next, passwords shorter than eight characters were excluded to align with modern minimum policy requirements, reducing the dataset by approximately 16.18 million entries and yielding a curated subset of 16.42 million passwords. The filtered RockYou dataset exhibits a median password length of nine characters $(M = 9.46, \sigma = 2.43)$. Notable outliers include lengthy HTML fragments or URLs used as passwords that likely reflect user behavior anomalies rather than deliberate choices. These entries were retained to preserve the dataset's authenticity despite their slight skewing effect on bit-strength calculations. Password entropy was estimated using zxcvbn's pattern-aware algorithm and showed a mean (median) entropy of 21.9 (21.7) bits with a standard deviation of 9.6 bits and 26.8 for the third quartile and 15.6 for the first quartile. Recognizing that RockYou reflects pre-2010 user behavior patterns, we constructed $D_{syn}$ by systematically doubling the bit-strength values of each password in the RockYou corpus while preserving its overall distribution shape. This approach accounts for improved password policies and heightened user awareness observed in recent years while maintaining compatibility with prior research methodologies. The synthetic dataset serves as an updated benchmark for evaluating Argon2's performance in higher-security contexts. The doubled values align with results from Komanduri et al. \cite{10.1145/1978942.1979321} showing that complex password requirements yield on average 44.67 bit-strength passwords.
\section{Attack Simulation Results}\label{sec:AttackSimResults}

\begin{figure}[ht]
    \centering
    \includegraphics[width=0.8\textwidth]{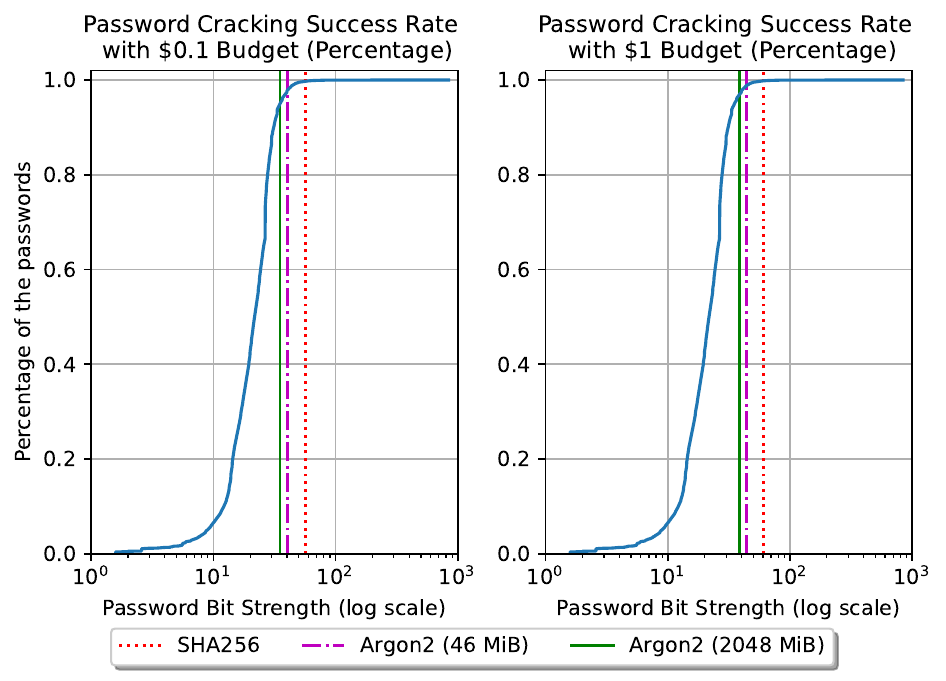}
    \caption{Password cracking success rates for the RockYou dataset under \$0.1 and \$1 budgets for SHA-256 and Argon2 configurations (46 MiB and 2048 MiB).}
    \label{fig:rockyou}
\end{figure}

\begin{figure}[ht]
    \centering
    \includegraphics[width=0.8\textwidth]{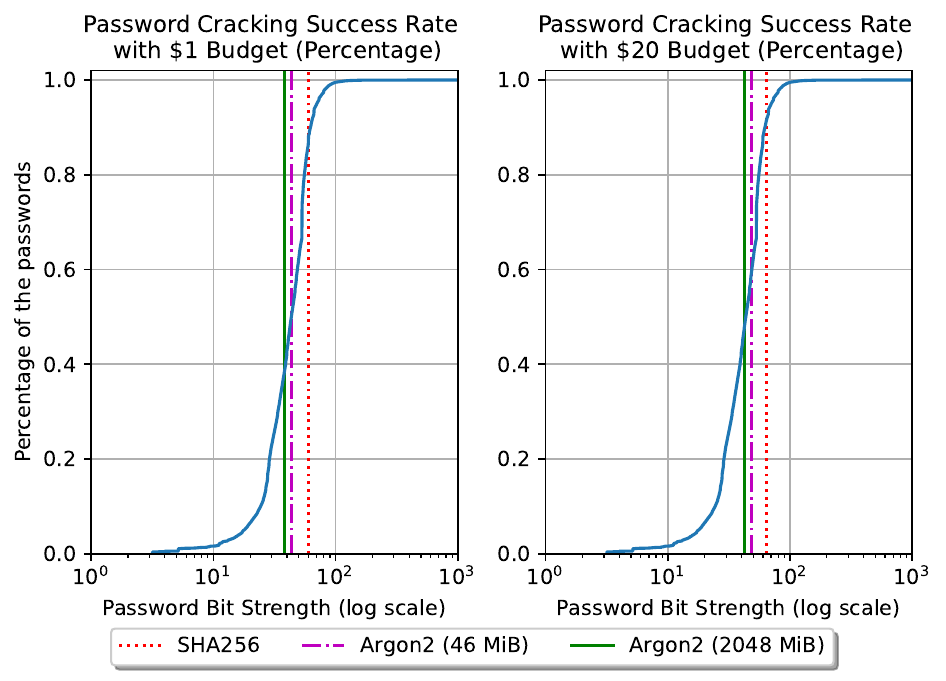}
    \caption{Password cracking success rates for the synthetic dataset (\(D_{\text{syn}}\)) under \$1 and \$20 budgets for SHA-256 and Argon2 configurations (46 MiB and 2048 MiB).}
    \label{fig:dsyn}
\end{figure}

Our attack simulations show fundamental security tradeoffs between hashing algorithms, parameter configurations, and password strength distributions. Figures \ref{fig:rockyou} and \ref{fig:dsyn} illustrate the compromise rates for SHA-256, Argon2 using OWASP's 46 MiB configuration, and Argon2 with RFC 9106's 2048 MiB configuration across varied attacker budgets for both datasets

\textbf{RockYou Dataset:} As depicted in Figure \ref{fig:rockyou}, SHA-256's susceptibility to low-cost attacks is strong, with near-total compromises (99.77\%) at just \$0.10 per account, slightly worsens at a \$1.00 budget (99.83\%). Conversely, Argon2 introduces modest resistance. The 46 MiB parameters reduce compromise rates to 98.81\% at \$1 budgets, while the 2048 MiB configuration cuts down success to 96.89\% under identical conditions, a notable 2.94\% improvement compared to SHA-256 protecting more than 475,000 accounts. While Argon2 impacts the attackers success to a limited extent, the weak passwords are the decisive factor. 

\textbf{Synthetic Dataset ($D_{syn}$)}: Figure \ref{fig:dsyn} shows a shift when modeling stronger user passwords and policies. Here, SHA-256 achieves higher resistance due to improved password bit-strength. Due to the exponential effects of password bit-strength, the gap between hashing algorithms widens. At \$1.00 budgets, 88.31\%, 50.74\% and 38.92\% of all passwords are compromised for SHA-256, Argon2 with 46 MiB and Argon2 with 2048 MiB. For the \$20 budget, this rate increases to 91.85\%, 59.16\% and 48.69\%. This demonstrates that the stronger 2048 MiB configuration does provide stronger protection compared to 46 MiB with a 11.82 (10.48) percentage points lower compromise rate under the \$1 (\$20) budget. However, the largest difference is the change from SHA-256 to the lower Argon2 configuration indicating that it provides significantly more protection. 

These results highlight the reliance on the strength of user passwords: For RockYou's median 21.7-bit passwords, even the strongest Argon2 configuration couldn't prevent attacker from cracking almost all passwords on small budgets. However, simulations using $D_{syn}$'s median 43.4-bit passwords show that Argon2 additionally protects 43.16\% of all accounts compared to SHA-256 with a \$20 budget. Our results show for attackers, that an increasing budget yields diminishing returns as the easy passwords are harvested fast while the increasing bitstrengths make attacks exponentially harder. At the same time, this also holds partially for defenders since strong parameter configurations do not help for weak user passwords. Using Argon2 instead of SHA-256 makes the biggest difference while the increasing server-side load does not proportionally protect more passwords but also shows diminishing returns for defenders. 
\section{Real-World Data Collection}\label{sec:data-collection}

Following our analysis of Argon2's theoretical security properties and its resilience under simulated attacks, we shift the focus to research questions 2–4. We now analyze Argon2's adoption trends through a systematic examination of GitHub repositories.

\subsection{Repository Search}\label{subsec:repo-search}
The dataset was constructed through systematic GitHub \cite{github2025} repository searches for five password hashing algorithms (Argon2, bcrypt, scrypt, yescrypt, and PBKDF2) across GitHub's entire availability period (2008-2024). Table \ref{tab:data_cleaning} shows the number of repositories for the filtering steps. For each algorithm, we executed temporal searches segmented by repository creation time, followed by a filtering phase. Effectively, we decided to include 31 repositories from one user for PBKDF2 and set the cutoff at 66 repositories and more per user. After manual review, we created a keyword list\footnote{\textit{miner}, \textit{mining}, \textit{proof-of-work}, \textit{proof of work}, \textit{currency}, \textit{coin}, \textit{wallet}, \textit{bitzeny}, \textit{doge}, \textit{mint}, \textit{blockchain}, \textit{contract}} related to common cryptocurrency themes to exclude irrelevant projects. After the filtering, we checked random samples from the results for each hashing algorithm and noticed that the 1,520 scrypt repositories still contained many repositories unrelated to password hashing but included similar terms (i.e. scrypto, bash scrypt, python scrypt), some of which we attribute to (intentional) misspellings. Therefore, we continued to filter the results and created an additional keyword list to ensure relevancy with words commonly used with hashing and key derivation functions: \textit{password, hash, auth, kdf, key derivation, percival} (the author of scrypt). 1,279 repositories of these 1,520 contained none of these additional keywords which we judged as too much. Therefore, we used an additional list containing similar names\footnote{\textit{scrypta}, \textit{scrypto}, \textit{scrypts}, \textit{scrypted}, \textit{scrypting}, \textit{inscryption}, \textit{scryptic}, \textit{scrypture}, \textit{ipsa-scrypt}} that only excluded 439 repos. Among these 439 repos, all but three contained none of the relevancy words and were subsequently excluded. The three outliers were manually reviewed and one marked as relevant resulting in remaining 1,081 hits ($1,520-439$). Out of these, only 238 contained at least one relevancy word, which led us to manually review the other results by analying the URL name and project description. We excluded projects if they were surely not related to password hashing, marked them as \textit{yes}, if we clearly connected them, and coded them as \textit{possible}, if we could not conclude with high certainty. The latter was the case for 326 repos that did not contain any description. This resulted in 595 repositories, including the 219 possible hits, identified as scrypt password hashing repositories. Including the possible hits rather overestimates the prevalence of scrypt password hashing than underestimating it. Since Argon2 was included in five repositories with a creation date before 2015 (the year it won the Password Hashing Competition), we reviewed them manually. Two of them included Argon2 later while the others are abandond or just include Argon2 references in non-productive parts.

\begin{table}[ht]
    \centering
    \caption{Repository collection and filtering statistics per algorithm}
    \label{tab:data_cleaning}
    \begin{tabular}{lllll}
    \toprule
    \textbf{Algorithm} & \textbf{Initial} & \textbf{Spam} & \textbf{Mining} & \textbf{Final} \\
                       & \textbf{Repos}  & \textbf{Removed (\%)} & \textbf{Filtered (\%)} & \textbf{Count} \\
    \midrule
    Argon2 & 1,602 & 534 (33.33\%) & 36 (2.25\%) & 1,032 \\  bcrypt & 12,727 & 604 (4.75\%) & 58 (0.46\%) & 12,065 \\
    scrypt & 2,396 & 528 (22.04\%) & 1,273 (51.13\%)* & 595 \\
    yescrypt & 76 & 0 (0\%)& 36 (47.37\%) & 40 \\
    PBKDF2 & 1,006 & 0 (0\%) & 12 (1.19\%) & 994 \\
    \bottomrule
    \end{tabular}
    \begin{flushleft}
        {\footnotesize * Includes extended relevance checks for scrypt repositories (see Subsection \ref{subsec:repo-search}).}
    \end{flushleft}
\end{table}
    
\vspace{-1em}

\subsection{Code Search}

The code search was performed using the GitHub Search API for programming languages supporting symbol extraction and additional programming languages that we assessed as relevant after our manual review of repositories, as listed in Subsection \ref{subsec:meth-parameter}. The primary search term for each query was the name of the password-hashing algorithm itself, refined with negative keywords to exclude cryptocurrency-related projects, which are outside the scope of this research. These negative keywords were the same as used in the repository search (see Subsection \ref{subsec:repo-search}). Furthermore, we excluded files with the \textit{.md} extension (to avoid \textit{README} files and other documentation) and files located within directories containing \textit{test} in their name to minimize irrelevant results. Due to the potential for a single repository to contain multiple instances of a given hashing function across different files, our search results often included duplicate entries for the same repository. To address this redundancy and estimate the number of unique repositories, we calculated a duplication quota based on the ratio of distinct repository IDs within the first 1,000 search results. This quota was then applied to the total number of search results to approximate the underlying number of unique repositories implementing each hashing algorithm. 

Table \ref{tab:code_search} shows the repository search results. Initial searches without programming language differentiation yielded total code hits of 48,768 for Argon2, 519,168 for bcrypt, 36,592 for PBKDF2, 131,328 for scrypt, and 3,232 for yescrypt.  Calculating repository redundancy required estimating a repository duplication quota based on unique repositories within the first 1,000 results (5.9\% for Argon2, 1.2\% for bcrypt, 13.6\% for PBKDF2, 26.2\% for scrypt, and 67.3\% for yescrypt). This resulted in estimated total repositories of 45,891, 512,938, 31,615, 96,920, 1,057, respectively. The search separated by programming languages and hash function has 115 combinations. For 52 combinations the query results were below 1,000 and the repository number could be counted directly without using the duplication quota. The overall results are shown in Table \ref{tab:code_search}. Since we could not conduct the additional filtering steps for scrypt that we did for the repository search, we used the filtering ratio from the repository search to estimate the filtered number of scrypt results. Table \ref{tab:redundancy-rates} shows the quotas for different result sizes and hashing functions.

\begin{table}
    \centering
    \caption{Total hits and estimated number of repos, separately for the different \\ password hashing methods and searches.}
    \label{tab:code_search}
    \begin{tabular}{lcccc@{}l@{}}
        \toprule
        & \textbf{Argon2} & \textbf{bcrypt} & \textbf{PBKDF2} & \textbf{scrypt} & \multicolumn{1}{c}{\textbf{yescrypt}} \\
        \midrule
        \multicolumn{6}{l}{\textit{Code Search with Programming Language Differentiation:}} \\
        \midrule
        Total Code Hits & 41,464 & 226,768 & 75,521 & 88,211 & 905 \\
        Estimated Total Repos & 33,170 & 213,012 & 64,645 & 64,416 & 531 \\
                              &        &         &        & (29,116*) &     \\
        \midrule
        \multicolumn{6}{l}{\textit{Simple Code Search:}} \\
        \midrule
        Total Code Hits  & 48,768 & 519,168 & 36,592 & 131,328 & 3,232 \\
        Estimated Total Repos & 45,891 & 512,938 & 31,615 & 96,920  & 1,057 \\
                                &         &         &         & (43,808*) &     \\
        \midrule
        \multicolumn{6}{l}{\textit{Repository Search Results:}} \\
        \midrule
        Results from Repo Search & 1,032 & 12,065 & 994 & 595 & 40 \\
        \bottomrule
    \end{tabular}
    \vspace{0.5em} 
    \begin{flushleft}
        {\footnotesize * After applying the scrypt false positive removal rate of 54.8\% as determined in \\ Subsection \ref{subsec:repo-search} after only applying the initial cryptocurrency filter.}
    \end{flushleft}
\end{table}

\begin{table}[htbp]
    \centering
    \caption{Repository redundancy quotas for password hashing methods from code search analysis with programming language differentiation. For large result sets (>3,500 hits, n=30) and medium-sized sets (1,000--3,500 hits, n=26), quotas were estimated from the first 1,000 results. For small result sets (<1,000 hits, n=52), exact quotas were determined. Analysis covers 108 of 115 programming language and hash function combinations.}
    \label{tab:redundancy-rates}
    \begin{tabular}{lcccccc}
        \toprule
        \textbf{Repo Redundancy} & \textbf{Argon2} & \textbf{bcrypt} & \textbf{PBKDF2} & \textbf{scrypt} & \textbf{yescrypt} & \textbf{Total} \\ 
        \midrule
        Estimated & 5.50\% & 9.89\% & 11.83\% & 24.03\% & --- & 13.67\% \\
        (>3500, n=30) & \small{n=3} & \small{n=12} & \small{n=7} & \small{n=8} & & \small{n=30} \\
        \addlinespace
        Estimated & 32.59\% & 23.60\% & 26.97\% & 44.15\% & --- & 32.36\% \\
        ($\leq3500$, n=26) & \small{n=9} & \small{n=4} & \small{n=7} & \small{n=6} & & \small{n=26} \\
        \addlinespace
        Exactly & 59.73\% & 51.17\% & 50.83\% & 64.11\% & 37.24\% & 50.87\% \\
        determined (n=52) & \small{n=11} & \small{n=7} & \small{n=9} & \small{n=9} & \small{n=16} & \small{n=52} \\
        \midrule
        Total & 42.03\% & 24.84\% & 31.70\% & 44.96\% & 25.90\% & 33.89\% \\
        (n=108*) & \small{n=23} & \small{n=23} & \small{n=23} & \small{n=23} & \small{n=16*} & \small{n=108*} \\
        \bottomrule
    \end{tabular}
    \vspace{0.5em} 
    \begin{flushleft} 
    {\footnotesize * In 7 of the total 115 cases, there were no matches for yescrypt and therefore not considered here.}
    \end{flushleft}
\end{table}

\subsection{Parametrisation}\label{subsec:parameterization}

For manual identification of Argon2 configurations, we decided to focus on high-quality repositories only. From the 1,068 Argon2 repositories found in the repository search, we focused on repositories with at least three star ratings, resulting in 253 remaining repositories. We further excluded repositories that were archived or not productive (e.g., described as homework assignments, demos or trials). We further divided the remaining 206 repositories into four equal sets based on their star count (3-4, 5-10, 11-30 and more than 30). We further excluded 21 repositories tied to cryptocurrency applications or password cracking. Afterwards, we classified each project in the following categories: components (libraries, wrapper, bindings), applications and sensitive applications (password manager and file encryption). Then we proceeded to manually extract the Argon2id configuration with a focus on iterations (\textit{t}) and memory (\textit{m}). In 24 repositories, we were unable to determine the parameters because they do not offer a (complete) default parametrisation, are specifications or benchmarking tools. If the software code used a library and didn't modify the parameter settings, we extracted the libraries default settings as parameters. In sum, parametrisation data was collected for 161 repositories.
\section{Real-World Implementations Analysis}\label{sec:repo-analysis}

\subsection{Adoption}

\begin{figure}[htbp]
    \centering
    \includegraphics[width=\textwidth]{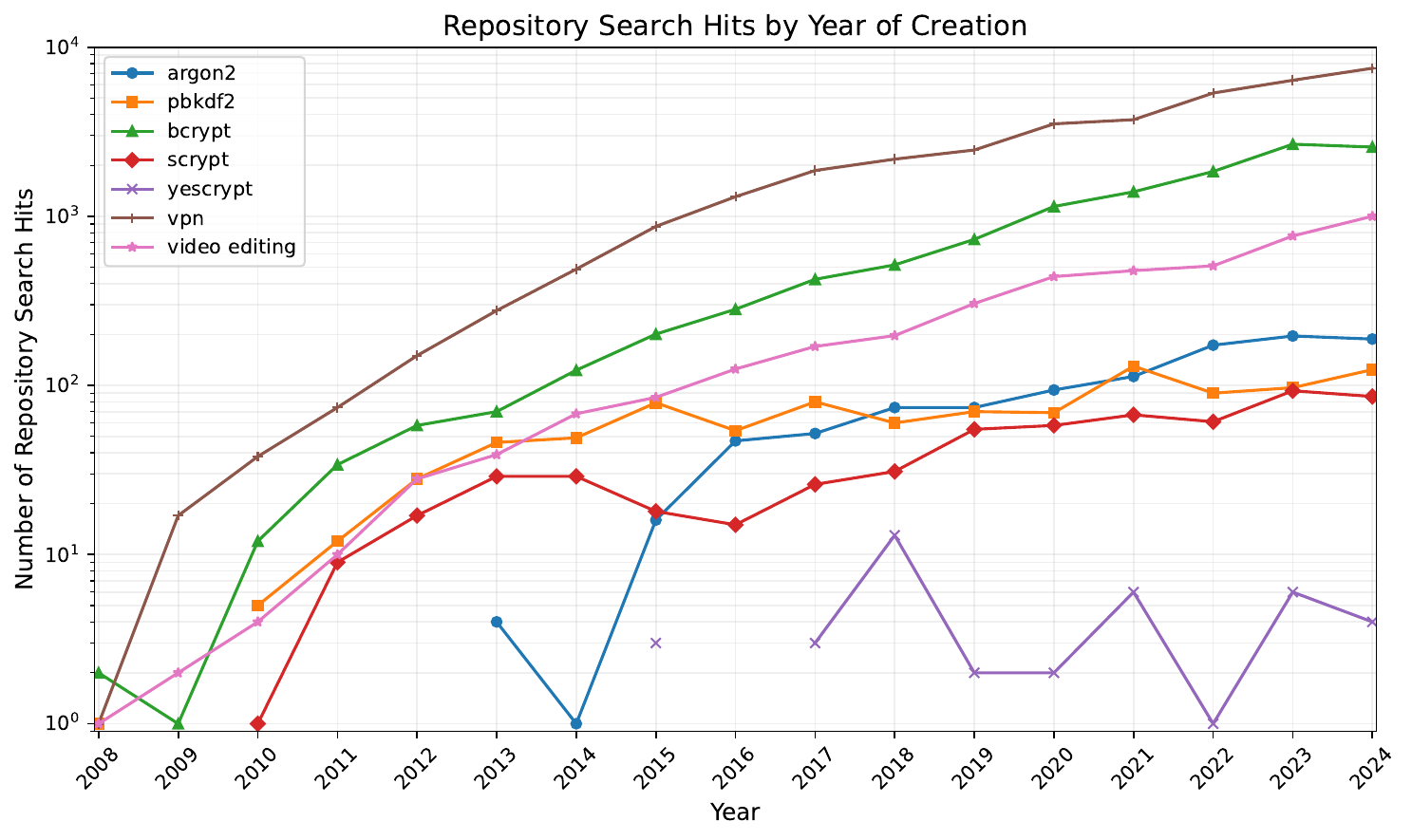}
    \caption{Repository search hits for various password hashing algorithms (2008-2024).}
    \label{fig:repo-search-results}
\end{figure}

Figure \ref{fig:repo-search-results} shows the number of repos for the hash functions per creation year. To put this into the context of general repository developments, we decided to introduce two more search terms, with \textit{VPN} being related to computer security and \textit{video editing} unrelated to the field. The development of bcrypt aligns with that of these additional search terms. Argon2 also shows continuous growth at a lower rate since its inception in 2015. scrypt and PBKDF2 show notably less development which we assume is due to the introduction of Argon2. Argon2 also managed to overtake the number of new repositories from 2018 onwards for scrypt, PBKDF2 and yescrypt. yescrypt, a competitor of Argon2 in the Password Hashing Competition, did not succeed in keeping up with Argon2's adoption and has stagnating lower creation numbers. Focusing on the time between 2015 and 2024, the average number of repositories with a creation date in the respective years has a mean value of 102.7 for Argon2 ($\sigma = 59.95$), 85.3 for PBKDF2 ($\sigma = 25.47$) and 51 for scrypt ($\sigma = 27.52$). We used the Kruskal-Wallis test to determine a statistically significant difference between the three groups ($H(2) = 6.07, p = .048$) and then conducted Dunn tests to compare the hashing functions with each other. Argon2 and scrypt differ significantly ($z = 2.13, p = .033$) and PBKDF2 and scrypt differ significantly ($z = 2.13, p = .033$) while there is no statistically significant difference between Argon2 and PBKDF2 ($z = 0, p = 1.0$). This indicates that Argon2 clearly overtook scrypt, but due to the larger variance in the number of Argon2 repositories created over the years no statistically significant overall difference is evident between Argon2 and PBKDF2.

During the repository analysis, we found that 141 repositories offered multiple hashing algorithms. The majority (93) implements exactly two functions (of the five in the scope of this analysis), 33 offer three functions and 15 offer 4 functions. From the possible 26 combinations, only 13 appear in our dataset. yescrypt appears only in two of these 13 combination repos and is therefore excluded from the following significance test. A chi-square goodness of fit test shows that there is no clear evidence that the distribution differs from uniform distribution ($\chi^2(3) = 0.25, p = .969$). Therefore, we did not find evidence that some of the analysed hashing functions are overrepresented in combination repositories.

The code search had more uncertainties since the number of repositories was estimated for search queries with more than 1,000 results based on the repo redundancy quota in the accessible results. Table \ref{tab:redundancy-rates} shows that the determined quotas clearly differ. Evidently, the estimated repository duplication quotas are closer to the exactly calculated ones when the number of code search results is smaller (below 3,500). This is plausible since the subset of 1,000 accessible code search results will be more representative if the selection pool is smaller (below 3,500 instead of more than 3,500). To ensure that the observed differences in repository duplication quotas are not systematically influenced by the interplay between hashing method and the size of the result set, we perform a chi-square test of independence, excluding yescrypt to avoid potential bias. The nonsignificant result ($\chi^2(6) = 8.36, p = .213$) supports the validity of comparing methods, indicating that the identified trends likely reflect genuine differences rather than sampling artifacts. Overall, the results from the general code search and the one segmented by programming languages supports the results from the repository search as shown in Table \ref{tab:code_search}. The repository search and simple code search results generally followed a consistent pattern. However, PBKDF2 was an exception to this trend, as the programming-segmented code search yielded more PBKDF2 results than the simple code search. We could not identify a plausible explanation for this anomaly. The other hash functions led to a smaller number of estimated repos in the segmented code search since the redundancy quota was more accurately calculated. The code search results support the repository search results for the adoption of Argon2, confirming its growing prominence among password hashing algorithms.

\subsection{Parametrization over Time}\label{subsec:param-time}

During the data collection phase, we used the star count as a quality proxy to focus on high quality repositories. We set the star count for at least 3 stars to conduct the initial filtering. To test our assumption, we divided this initial set of 253 repositories into four equal size classes: 3-4 stars, 5-10 stars, 11-30 stars and more than 30 stars. We created a contingency table with the star sets and the number of productive and non-productive repositories that we filtered in the following step. A chi-square independence test shows the distribution is not independent ($\chi^2(6) = 12.53, p = .006$) and that higher star counts are associated with less non-productive repositories. This supports our assumption of the star count as a quality proxy. After extracting the parameter configurations successfully from the selected 161 repositories (see \ref{subsec:parameterization}), we noticed that the configurations were clustered with the following most popular configurations:

\begin{itemize}
	\item $t=3, m=4096$ KiB (33 times) 
	\item $t=3, m=65536$ KiB (28 times),
	\item $t=2, m=19456$ KiB (11 times),
	\item $t=1, m=65536$ KiB (10 times),
	\item $t=2, m=65536$ KiB (9 times)
\end{itemize}

\begin{figure}[htbp]
    \centering
    \includegraphics[width=\textwidth]{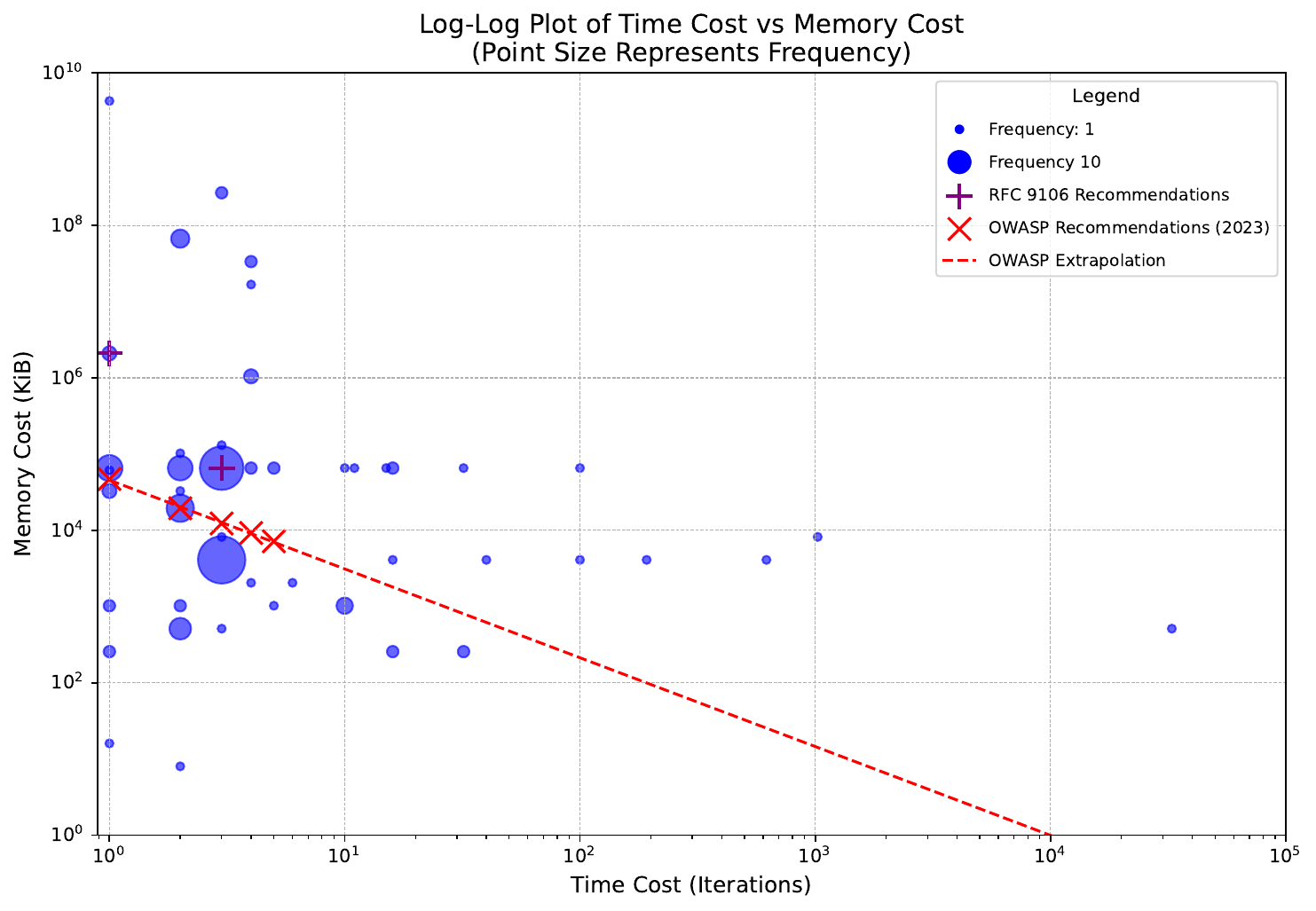} 
    \caption{
        Log-log plot of time cost (\textit{t}) versus memory cost (\textit{m}). 
        The size of each blue dot represents the frequency of data points with a specific combination of \textit{t} and \textit{m}. 
        The red crosses represent OWASP recommendations, while the purple crosses represent RFC 9106 recommendations. 
        The dashed red line extrapolates OWASP recommendations using a linear regression in log-log space.
            }
    \label{fig:log_log_plot}
\end{figure}

We attribute this especially to the default values of used libraries. We compared the parameter strength by linearly extrapolating the OWASP Argon2 recommendations and classifying the extracted configurations as weaker or stronger. Figure \ref{fig:log_log_plot} shows the OWASP extrapolation and the observed respository configurations. 75 repositories were weaker and 86 stronger than the OWASP recommendations. To see the development over the years, we created a contingency table shown in Table \ref{tab:repo_age_param_strength} and grouped the repository creation years 2013 - 2024 in three groups: before 2018, 2019-2021, 2022-2024. The grouping helped to reach the minimum number of entries for the chi-square test and takes the publication date of RFC 9106 and OWASP recommendations into account. A chi-square independence test confirms that these factors are not independent ($\chi^2(2) = 8.42, p = .015$). This shows an increase of parametrization strength over time.

\begin{table}[ht]
    \centering
    \caption{Repository age and categorical strength of parameterization.}
    \label{tab:repo_age_param_strength}
    \begin{tabular}{@{}lcccc@{}}
    \toprule
     & \multicolumn{3}{c}{\textbf{Repository Age}} & \\
    \cmidrule(lr){2-4}
    \textbf{Parameterization} & \textbf{$\leq$2018} & \textbf{2019–2021} & \textbf{2022–2024} & \textbf{Total} \\
    \textbf{Strength} & & & & \\ \midrule
    Weaker & 38 (60.3\%) & 23 (41.1\%) & 14 (33.3\%) & 75 (46.6\%) \\
    Stronger & 25 (39.7\%) & 33 (58.9\%) & 28 (66.7\%) & 86 (53.4\%) \\ \midrule
    Total & 63 (100.0\%) & 56 (100.0\%) & 42 (100.0\%) & 161 (100.0\%) \\ \bottomrule
    \end{tabular}
\end{table}

\subsection{Parametrization by Program Type}

Table \ref{tab:parameterization} shows the parametrization strength for the different software types. Weaker configurations are more present for components while the configurations are stronger for applications. A chi-square test confirms this hypothesis by showing a statistically significant effect of the software type variable ($\chi^2(2) = 7.38, p = .007$). When the applications are separated into sensitive (file encryption, password managers) and normal password hashing applications, there is no statistically significant effect ($\chi^2(1) = 0.002, p = .967$) between them. So, the hypothesis that sensitive software programs use stronger parametrizations could not be confirmed. To exclude effects from the star count on the software type, we tested with a contingency table if there is a statistically significant effect of the repository star count (divided in four equal sized groups) on the software type (application vs. component) and determined no significant effect ($\chi^2(3) = 3.41, p = .332$). 

\begin{table}[ht]
    \centering
    \caption{Cross table: Software type x categorical strength of parameterization.}
    \label{tab:parameterization}
    \begin{tabular}{lcccc}
        \toprule
        \textbf{Parameterizations} & \makecell{\textbf{Sensitive}\\\textbf{Application}} & \textbf{Application} & \textbf{Component} & \textbf{Total} \\
        \midrule
        Weaker & 4 (26.7\%) & 6 (27.3\%) & 65 (52.4\%) & 75 (46.6\%) \\
        Stronger & 11 (73.3\%) & 16 (72.7\%) & 59 (47.6\%) & 86 (53.4\%) \\
        \midrule
        Total & 15 (100.0\%) & 22 (100.0\%) & 124 (100.0\%) & 161 (100.0\%) \\
        \bottomrule
    \end{tabular}
\end{table}

Additionally, we analysed how the parametrization is affected by the star count. Surprisingly, stronger configurations dominate the repositories with lower star counts (3-10), while repositories with more stars (more popular repositories) exhibit weaker configurations (see Table \ref{tab:star_categories_param_strength}). A chi-square test confirmed this effect as statistically significant ($\chi^2(3) = 8.71, p = .033$). While this seems counterintuitive, we noticed that the star count (popularity) is tied to the age as well. This hypothesis would fit to the finding from Subsection \ref{subsec:param-time} that older repositories exhibit weaker configurations than newer ones. Therefore, we tested the influence of repository age on star count and confirmed with the chi-square independence test that older repositories have a higher star count ($\chi^2(6) = 32.53, p < .001$). This is also intuitive since older projects have more time to gain popularity and accrue stars. Finally, we tested the influence of repository age on software type and found that there is a statistically significant effect such that the number of components is higher for older repositories while the younger repositories include more applications ($\chi^2(2) = 20.25, p < .001$). This weakens the finding that components exhibit weaker configurations since the repository age is a mitigating factor here as well, suggesting that the observed difference may be more strongly associated with when repositories were created rather than their functional purpose as components versus applications.

\begin{table}[ht]
    \centering
    \caption{Star categories and categorical strength of parameterization.}
    \label{tab:star_categories_param_strength}
    \begin{tabular}{@{}lccccc@{}}
    \toprule
     & \multicolumn{4}{c}{\textbf{Repository Star Count}} & \\
    \cmidrule(lr){2-5}
    \textbf{Parameterization} & \textbf{3–4 Stars} & \textbf{5–10 Stars} & \textbf{11–30 Stars} & \textbf{>30 Stars} & \textbf{Total} \\
    \textbf{Strength} & & & & & \\ \midrule
    Weaker & 13 (34.2\%) & 12 (35.3\%) & 28 (62.2\%) & 22 (50.0\%) & 75 (46.6\%) \\
    Stronger & 25 (65.8\%) & 22 (64.7\%) & 17 (37.8\%) & 22 (50.0\%) & 86 (53.4\%) \\ \midrule
    Total & 38 (100.0\%) & 34 (100.0\%) & 45 (100.0\%) & 44 (100.0\%) & 161 (100.0\%) \\ \bottomrule
    \end{tabular}
\end{table}

\section{Discussion}\label{sec:Discussion}

\subsection{Main Findings}

\textbf{Argon2's security advantages over SHA-256} Our attack simulations validate Argon2's fundamental security advantage over SHA-256, even with modest parameters. For synthetic datasets modeling modern password policies ($D_{\text{syn}}$), Argon2 (46 MiB) reduces compromise rates by 37.57\% versus SHA-256 at \$1 account budgets. However, memory allocation effectiveness exhibits diminishing returns: while 2048 MiB configurations provided  23.3\% (at \$1) and 17.7\% (at \$20) additional protection over 46 MiB, their 44.5\texttimes greater memory demands impose impractical scaling costs for many systems. Crucially, no configuration sufficiently mitigates risks for weak passwords, emphasizing that algorithm selection cannot compensate for poor user credential practices.

\textbf{Growing adoption trend} Argon2 adoption has steadily increased since its introduction in 2015, surpassing competing algorithms like scrypt and PBKDF2 in the number of new GitHub repositories created annually starting in 2018. However, its adoption lags behind bcrypt, which remains the most widely implemented password hashing algorithm likely due its older age and familiarity of developers. Our analysis of over 161 manually reviewed repositories revealed that Argon2 is present in a diverse range of software projects, with both applications (38 repositories) and components (124 repositories) incorporating it into their cryptographic workflows. Interestingly, the frequency of Argon2's coexistence with other algorithms (e.g., bcrypt, PBKDF2, scrypt) in multi-algorithm implementations underscores its growing acceptance as part of a broader cryptographic toolbox. However, the absence of statistically significant overrepresentation of any specific combinations suggests that Argon2 adoption is not yet widespread enough to dominate as a preferred choice.

\textbf{Parameter evolution} Our analysis of Argon2 parameter configurations in real-world implementations reveals a shift towards stronger configurations over time. Before 2018, 60.3\% of repositories adopted weaker-than-OWASP-recommended settings, but this proportion decreased to 33.3\% in repositories created after 2022. This trend aligns with the publication of the RFC 9106 standard in 2021 and evolving OWASP guidelines, which have likely increased awareness of the importance of using secure configurations. The observed clustering of common configurations suggests a heavy reliance on default library settings rather than deliberate customization by developers.

\textbf{Context-depending Configuration} Contrary to our expectations, sensitive applications such as password managers and file encryption software did not consistently implement stronger Argon2 parameter configurations compared to general-purpose applications. While 73.3\% of sensitive applications used stronger settings than OWASP recommendations, this proportion was similar to general application repositories (72.7\%). This finding highlights unclear practices for parameter selection, even among software with higher security stakes. Interestingly, components (e.g., libraries and cryptographic bindings) exhibited weaker parameterization (52.4\% below OWASP standards), possibly due to the need to balance performance and usability across diverse deployments. It is worth noting that this correlates with older repositories implementing weaker configurations on average and younger repositories increasingly being applications. Similarly, repositories with a higher star count tend to implement weaker configurations which is likely also connected to repository age.

\subsection{Practical Recommendations}

Our comprehensive analysis reveals that while Argon2 offers theoretical security advantages, these are not fully realized in practice. To bridge this gap, we offer actionable recommendations. Argon2's benefits are amplified when combined with robust user passwords. Studies indicate that password meters and password policies significantly increase password strength, aiding users in creating more secure credentials \cite{10.1145/1978942.1979321,10.5555/2362793.2362798}. Developers should implement strength estimation tools, enforce password policies, and integrate with techniques like password blacklisting to ensure users generate stronger passwords. Facilitating secure implementation requires simplifying the developer experience. Integrating comprehensive documentation and automated parameter selection tools into cryptographic libraries can increase adoption \cite{DBLP:conf/sp/Acar0FGKMS17}. Providing sane default configurations (OWASP and RFC 9106 recommendations) directly within libraries streamlines the process, reducing configuration errors. Additionally, automated checks in vulnerability scanners or compilers could flag weak settings, ensuring password-hash hardening remains a priority. Argon2 parameters may create a configuration challenge for some developers. Implementing adaptive benchmarking tools, as noted in some repositories, automates parameter selection tailored to local environments. To protect servers from potential denial-of-service attacks resulting from intensive hash computation, partial client-side hashing can be considered.  
Clients precompute their passwords with Argon2 before transmitting the resulting hash to the server, which then performs fast hashing functions \cite{8406611}. While OWASP suggests a conservative 46 MiB of memory on the client side, current devices often possess capabilities to accommodate higher allocations, allowing for more robust defense multipliers without compromising usability.

\subsection{Limitation and Future Work}

The economic models for hash computation costs use cryptocurrency mining practices as a proxy for large-scale computational attacks. While this provides a validated cost structure, it does not fully capture the nuances of different attackers that can hardly compete with large centralized mining pools. Using cryptocurrency mining as a proxy for attacker costs introduces uncertainty into our budget-scenario analyses, which could systematically underestimate the true protection levels afforded by Argon2 in production environments. Also, the used budget may vary significantly: While \$1 might be appropriate for a low-relevance credentials, accounts with cryptocurrency assets carry significantly more wealth that in turn justifies substantially increased attack budgets. The password datasets we employed (RockYou and $D_{syn}$) serve as benchmarks for password strength but carry inherent limitations. RockYou is decades old and its password distribution might not accurately reflect current practices. While our synthetic dataset addresses this by doubling bit-strength values, it remains an approximation. Real-world password behavior, influenced by contemporary policies, user awareness, and cultural factors, might deviate significantly from our modeled datasets.

Our real-world study focuses on open-source projects hosted on GitHub. This selection bias may skew our results towards projects developed in a particular community culture, potentially missing trends in proprietary software or repositories on alternative platforms. The reliance on GitHub's prominence means that our findings might underestimate the adoption and implementation trends of Argon2 in closed-source or enterprise environments, where different regulations and development practices could influence cryptographic choices. Moreover, developer motivations for starring repositories are heterogeneous (e.g., bookmarking, acknowledgment of quality, personal interest), adding variability to our quality proxy metric. Additionally, the manual extraction process restricted our parameter analysis to 161 high-star repositories, potentially missing patterns present in a broader sample. While we believe this selection provides a representative set of quality implementations, broader sampling might reveal different distributions in parameter configurations. Addressing these limitations requires further research, potentially incorporating a broader range of software implementations, cost models derived from real-world attackers and conducting user studies with programmers to explore Argon2 transition barriers.
\section{Conclusion}\label{sec:Conclusion}

This research evaluated Argon2's cryptographic effectiveness for password hashing and relevant implementation trends in real-world software environments. Through attack simulations we demonstrated that Argon2 provides substantial security advantages over SHA-256, with the 2048 MiB RFC 9106 configuration reducing compromise rates by 46.99\% compared to SHA-256 at \$20 attack budgets in datasets modeling modern password policies. The OWASP-recommended configuration offers less protection for robust user passwords compared to RFC 9106 values. However, even the strongest Argon2 configurations cannot compensate for weak user passwords, demonstrating the need for robust user passwords. Beyond technical performance, the analysis of GitHub repositories revealed that real-world adoption of Argon2 has grown steadily, surpassing other modern algorithms like scrypt and PBKDF2 in new implementations since 2018. Despite this, bcrypt remains the dominant choice, and parameter configurations in many repositories still fall short of OWASP and RFC 9106 recommendations. Moreover, parameter strength has improved over time, aligning with updated standards and adoption, but a significant number of implementations have weaker-than-recommended configurations. Sensitive applications do not consistently implement stronger Argon2 configurations compared to others, challenging assumptions about the correlation between software security demands and cryptographic diligence. These results suggest that while Argon2 holds significant cryptographic advantages for password hashing, its real-world security effectiveness depends on proper parameterization and user practices.

\begin{credits}

\subsubsection{\discintname}
The authors have no competing interests to declare that are relevant to the content of this article.
\end{credits}

\bibliographystyle{splncs04}
\bibliography{cite}

\end{document}